\documentclass[fleqn,usenatbib,useAMS]{mnras}
\usepackage{mathtools}
\usepackage{txfonts}

\usepackage[T1]{fontenc}
\usepackage{ae,aecompl}

\usepackage{color}
\usepackage{graphicx}

\usepackage{amsmath,amssymb,amsfonts}

\def\mns{M}
\def\eq{Eq.}
\def\eqs{Eqs.}
\def\prl{Phys. Rev. Lett.}
\def\prd{Phys. Rev. D}

\def\apj{Astrophys. J.}

\def\aap{Astronomy and Astrophysics}

\def\mnras{Mon. Not. R. Astron. Soc.}

\def\apss{Astrophys. Space Sci.}

\newcommand{\oldtext}[1]{}

\title[Bondi accretion for stiff equations of state]{Relativistic Bondi accretion for stiff equations of state}

\author[Richards, Baumgarte \& Shapiro]{Chloe B.\ Richards,$^{1}$ Thomas W.\ Baumgarte,$^{1}$ and Stuart L.\ Shapiro$^{2,3}$ \\
$^{1}$Department of Physics and Astronomy, Bowdoin College,
 Brunswick, ME 04011 \\
$^{2}$Department of Physics, University of Illinois at Urbana-Champaign, Urbana, IL 61801\\
$^{3}$Department of Astronomy and NCSA, University of Illinois at Urbana-Champaign, Urbana, IL 61801}

\date{Accepted XXX. Received YYY; in original form ZZZ}

\pubyear{2021}

\begin{document}
\label{firstpage}
\pagerange{\pageref{firstpage}--\pageref{lastpage}}
\maketitle

\begin{abstract}
We revisit Bondi accretion -- steady-state, adiabatic, spherical gas flow onto a Schwarzschild black hole at rest in an asymptotically homogeneous medium -- for stiff polytropic equations of state (EOSs) with adiabatic indices $\Gamma > 5/3$.  A general relativistic treatment is required to determine their accretion rates, for which we provide exact expressions.  We discuss several qualitative differences between results for soft and stiff EOSs -- including the appearance of a minimum steady-state accretion rate for EOSs with $\Gamma \geq 5/3$ -- and explore limiting cases in order to examine these differences.  As an example we highlight results for  $\Gamma = 2$, which is often used in numerical simulations to model the EOS of neutron stars. We also discuss a special case with this index, the ultra-relativistic `causal' EOS, $P = \rho$.  The latter serves as a useful limit for the still undetermined neutron-star EOS above nuclear density. The results are useful, for example, to estimate the accretion rate onto a mini-black hole residing at the center of a neutron star.
\end{abstract}


\section{Introduction}

\cite{Bon52} accretion describes the steady-state, spherically symmetric, adiabatic flow of gas onto a 
point mass in Newtonian gravitation. Far from the point mass it is assumed that the gas is infinite in extent, homogeneous and at rest, and that its self-gravity can be ignored. While these idealized assumptions rarely will be realized in nature, Bondi flow nevertheless captures many of the qualitative properties of some accretion flows and provides reasonable estimates of the accretion rates onto stars; it therefore has been invoked frequently to approximate astrophysical accretion processes.  In addition, relativistic Bondi accretion flow provides a powerful test for numerical relativity codes designed to handle relativistic hydrodynamics in the presence of black holes.

For isentropic fluids with adiabatic indices 
$1 \leq \Gamma \leq 5/3$, and pressure laws of the form $P=K\rho_0^{\Gamma}$, where $\rho_0$ is the rest-mass density and $K$ is a constant, the solutions admit a critical point where the flow becomes transonic and the accretion rate assumes its maximum value. This maximal rate can be obtained analytically (see, e.g., Chapter 14 in \cite{Shapiro}, hereafter ST, for a textbook treatment). For $\Gamma = 5/3$ the critical point lies at the origin in a Newtonian treatment. For stiffer equations of state, with $\Gamma > 5/3$, the Newtonian approach yields unphysical solutions, so that this case requires a fully relativistic treatment. While $1 \leq \Gamma \leq 5/3$ is suitable for many astrophysical accretion processes, accretion of gas obeying a stiffer equation of state (EOS) with higher $\Gamma$ has recently attracted some increased attention, since it arises, for example, in the hypothetical scenario of a small, possibly primordial, black hole residing at the center of a neutron star \citep[see, e.g.,][and references therein]{CapPT13,EasL19,GenST20}. While accretion rates onto a black hole at the center of main-sequence stars can be approximated by Bondi values for gas with $\Gamma \leq 5/3$ and asymptotic sound speeds much less than the speed of light \citep[e.g.][]{Mar95}, both the accretion rate onto and the flow parameters near a black hole at the center of a neutron star require the relativistic equations that apply to stiffer equations of state and strong gravitational fields.

The general relativistic analogue of the Bondi equations was first derived by \cite{Mic72} for adiabatic accretion onto a (nonrotating) Schwarzschild black hole.  ST proved that for any EOS obeying the causality constraint that the speed of sound is less than the speed of light,
the flow {\it must} pass through a critical point outside the event horizon, even for $\Gamma = 5/3$. 
Most importantly, their result immediately implies that for {\it any} $\Gamma$, the maximal accretion rate and corresponding critical flow solution is the {\it unique} steady-state solution for spherical, adiabatic accretion onto a Schwarzschild black hole. The alternative subsonic flow solutions with smaller accretion rates allowed by the Newtonian Bondi equations, let alone the limiting case of zero accretion for hydrostatic atmospheres, are ruled out for black holes.

While Bondi accretion for stiff equations of state, with $\Gamma > 5/3$, has been discussed by several authors, most of their treatments invoked mathematical approximations \citep[see, e.g.,][]{Beg78}, or focused on the number and nature of transonic points \citep[e.g.][]{Ray80,Cha85,Das02} or on the existence and formal mathematical properties of solutions \citep[e.g.][]{ChaMS16}. Our paper complements these previous studies, as well as the corresponding treatments in textbooks, in several ways.  We provide exact expressions for the dimensionless accretion eigenvalues $\lambda_{\rm GR}$ (see \eq~\ref{lambda_GR} and Fig.~\ref{fig:lambdas} below) that determine the (critical) accretion rates (\ref{Mdot}).  We discuss several qualitative differences between accretion of fluids with soft ($\Gamma < 5/3$) and stiff ($\Gamma > 5/3$) EOSs, including the appearance of a minimum steady-state accretion rate (see \eq~\ref{Mdot_min}) and an asymptotic power-law density profile (see \eq~\ref{powerlaw_profile}) for stiff EOSs.  We explore these differences by considering, as limiting cases, both low and high-density regimes.  We highlight, as an illustrative example of particular relevance for the interior of neutron stars, results for $\Gamma = 2$, which includes the ultra-relativistic `causal' EOS $P = \rho$ as a limiting case.  Here $P$ is the gas pressure and $\rho$ the total mass-energy density. We adopt geometrized units with $G \equiv 1 \equiv c$ throughout this paper.

\section{Exact treatment}

Following \cite{Mic72} (see also Appendix G in ST for a textbook treatment) we assume that the fluid's self-gravity can be ignored, so that, further assuming spherical symmetry, the spacetime metric is given by the Schwarzschild metric
\begin{equation}
    ds^2 = - \left( 1 - \frac{2M}{r}\right) \, dt^2 + \left( 1 - \frac{2M}{r} \right)^{-1} dr^2 + r^2 d \Omega^2.
\end{equation}
Here we have adopted Schwarzschild coordinates, and $M$ is the black hole's mass.  The accretion rate $\dot M$ can then be expressed in terms of fluid values at the critical areal radius $r_s$,
\begin{equation}\label{M_dot_ana}
\dot{M} = 4\pi{\rho}_{0s} {u_s} r_s^2,
\end{equation}
were $\rho_{0}= m n$ is the rest-mass density, with $m$ the mean baryon mass and $n$ the baryon number density, and $u$ is the negative radial component of the fluid 4-velocity, $u \equiv -u^r$, measuring the {\em inward} flow.  The rest-mass density and the four-velocity have to satisfy the relativistic Euler equation and the equation of baryon conservation, which we combine and list the results, for completeness, as \eq~(\ref{prime}) in Appendix \ref{sec:profiles}.  The critical radius $r_s$, which coincides with the transonic radius for nonrelativistic flows, is defined by a simultaneous vanishing of the numerators and denominators (\ref{flow_coefficients}) appearing in these equations, which results in \eqs~(\ref{u_s}) and (\ref{r_s}) below.  We further assume that $u \rightarrow 0$ as $r \rightarrow 0$, and that the flow is subsonic asymptotically, in which case the flow has to pass through a critical point (see Appendix G in ST).  Throughout this paper the subscript $s$ denotes the value of any variable at the critical radius, and $\infty$ will denote its value in the asymptotic region $r \rightarrow \infty$.

Enforcing the flow at the critical radius to remain regular results in the conditions
\begin{equation}\label{u_s}
u_s = \frac{a_s}{a_\infty} \left(1 + 3a_s^2\right)^{-1/2} \, a_\infty
\end{equation}
and
\begin{equation} \label{r_s}
r_s = \frac{a_\infty^2}{a_s^2} \frac{1 + 3 a_s^2 }{2} \, \frac{M}{a_\infty^2}
\end{equation}
(see \eq~G.17 in ST, hereafter ST G.17).  Here $a$ is the sound speed, related to the total mass-energy density $\rho$ and the pressure $P$ by
\begin{equation} \label{soundspeed}
a^2 = \left. \frac{dP}{d\rho} \right|_s
= \left. \frac{dP}{d \rho_0} \right|_s \frac{\rho_0}{\rho + P},
\end{equation}
where derivatives are taken at constant entropy.  Assuming a polytropic EOS 
\begin{equation} \label{polytrope}
P = K \rho_0^\Gamma,
\end{equation}
where $K$ is a constant and $\Gamma$ the adiabatic index, the sound speed is related to rest-mass density by
\begin{equation} \label{a2_of_rho0}
    a^2 = \frac{\Gamma K \rho_0^{\Gamma - 1}}{1 + \Gamma K \rho_0^{\Gamma - 1}/(\Gamma - 1)}.
\end{equation}
Given values of $\rho_0$ and $a$, for example in the asymptotic region, the constant $K$ in (\ref{polytrope}) can be found from
\begin{equation} \label{K}
    K = \frac{1}{\Gamma \rho_0^{\Gamma - 1}} \,
    \frac{a^2}{1 - a^2/(\Gamma - 1)}.
\end{equation}
For ideal, degenerate Fermi gases, $K$ can be expressed in terms of constants of nature (see, e.g., ST 2.3.22 and ST 2.3.23).  In the following we will assume only that $K$ is some positive constant.  Note that only sound speeds $a < a_{\rm max}$ with
\begin{equation} \label{amax}
    a^2_{\rm max} = \Gamma - 1
\end{equation}
correspond to finite rest-mass densities.  We observe from \eq~(\ref{u_s}) that the critical point is {\em not} a transonic point in the relativistic solution, since $u_s < a_s$.

Evaluating \eq~(\ref{a2_of_rho0}) both at $r_s$ and in the asymptotic region yields a relation between $\rho_{0s}$ and the rest-mass density's asymptotic value $\rho_{0\infty}$,
\begin{equation}\label{rho_0}
\rho_{0s} = \left(\frac{a_s}{a_\infty}\right)^{2/(\Gamma - 1)} \left( \frac{\Gamma - 1 - a_\infty^2}{\Gamma  - 1 - a_s^2} \right)^{1/(\Gamma - 1)} \, \rho_{0\infty}.
\end{equation}
Inserting (\ref{u_s}), (\ref{r_s}), and (\ref{rho_0}) into (\ref{M_dot_ana}) we now find
\begin{equation} \label{Mdot}
\dot M = 4 \pi \lambda_{\rm GR} \left( \frac{M}{a_\infty^2} \right)^2 \rho_{0\infty} a_\infty,
\end{equation}
where
\begin{equation} \label{lambda_GR}
\lambda_{\rm GR} \equiv \left( \frac{a_s}{a_\infty} \right)^{(5 - 3 \Gamma)/(\Gamma - 1)} 
\left( \frac{\Gamma - 1 - a_\infty^2}{\Gamma  - 1 - a_s^2} \right)^{1/(\Gamma - 1)} \frac{(1 + 3 a_s^2)^{3/2}}{4}
\end{equation}
is a dimensionless ``accretion rate eigenvalue".

In the Newtonian limit, with $a_s \ll 1$ and $a_\infty \ll 1$, the relativistic expression (\ref{lambda_GR}) reduces to the Newtonian value $\lambda_s$,
\begin{equation} \label{lambda_N}
\lambda_{s} = \frac{1}{4} \, \left( \frac{a_s}{a_\infty} \right)^{(5 - 3 \Gamma)/(\Gamma - 1)}.
\hfill
\mbox{~~~~~(Newtonian)}
\end{equation}
From \eq~(\ref{as_soft}) below, or (ST 14.3.14), we also have
\begin{equation} \label{as_o_ainfty}
\frac{a_s}{a_\infty} = \left( \frac{2}{5 - 3 \Gamma} \right)^{1/2}     \hfill
\mbox{~~~~~(Newtonian)}
\end{equation}
in the Newtonian limit, which we can insert into (\ref{lambda_N}) to obtain the Newtonian accretion eigenvalues
\begin{equation} \label{lambda_s}
\lambda_s = \frac{1}{4} \left( \frac{2}{5 - 3 \Gamma} \right)^{(5 - 3 \Gamma)/2 (\Gamma - 1)}\hfill
\mbox{~~~~~(Newtonian)}
\end{equation}
(compare ST 14.3.17 and Table ST 14.1).  Evidently, the Newtonian treatment breaks down for $\Gamma > 5/3$.  Note also that the Newtonian eigenvalues $\lambda_s$ depend on $\Gamma$ only, while the relativistic values $\lambda_{\rm GR}$ also depend on the critical and asymptotic values of the sound speed, $a_s$ and $a_\infty$.

We can relate $a_s$ and $a_\infty$ by evaluating the integrated relativistic Euler equation for stationary flow at the critical radius $r_s$, 
\begin{equation}\label{setup_cubic}
(1 + 3a_s^2)\left(1 - \frac{a_s^2}{\Gamma - 1}\right)^2 =\left(1 - \frac{a_\infty^2}{\Gamma - 1}\right)^2
\end{equation}
(see ST G.30), which we now write as a cubic equation for $a_s^2$
\begin{equation}\label{our_cubic}
    a_s^6 + \frac{a_s^4}{3}(7 - 6\Gamma) + \frac{a_s^2}{3}(3\Gamma^2 -  8\Gamma + 5) + \frac{a_\infty^2}{3}(2\Gamma - 2 -a_\infty^2) = 0,
\end{equation}
\citep[cf.][]{Beg78}. We write this equation in the form
\begin{equation}\label{gen_cubic}
    x^3 + Ax^2 + Bx + C = 0
\end{equation}
for $x = a_s^2$ and identify the coefficients
\begin{equation}\label{cubic_components}
    \begin{split}
    A & = \frac{1}{3} \, (7-6\Gamma) \\
    B & = \frac{1}{3} \, (1 - \Gamma) \,(5 - 3\Gamma)     \\
    C & = \frac{a_\infty^2}{3} \, (2\Gamma - 2 -a_\infty^2),
    \end{split}
\end{equation}
noting that all three coefficients $A$, $B$, and $C$ are real.

As an aside we note that (\ref{setup_cubic}) can also be written as a quadratic equation for $a_\infty^2$, which is then solved by
\begin{equation} \label{gen_quadratic}
    a_\infty^2 = \Gamma - 1 \pm \left\{
    \left(\Gamma - 1\right)^2 + 3 \left(a_s^6 + A a_s^4 + B a_s^2 \right)
    \right\}^{1/2}.
\end{equation}
For most applications, however, it is more useful to consider $a_\infty$ as given, which then requires solving the cubic equation (\ref{gen_cubic}) for $a_s$.

Before proceeding we observe a remarkable property of \eq~(\ref{gen_cubic}).  Specifically, the equation allows a double root, meaning that it can be written in the form
\begin{equation} \label{double_root_factors}
    (x - x_1)^2 (x - x_2) = 0,
\end{equation}
if $a_\infty^2 = \Gamma - 1$, in which case $x_1 = \Gamma - 1$ also (see Appendix \ref{sec:double_roots}).  Therefore, sequences of physically viable solutions, parameterized by $a_\infty$ and satisfying the constraint $a < a_{\rm max} = \Gamma - 1$ (see \eq~\ref{amax}), terminate at a double root of \eq~(\ref{gen_cubic}) with 
\begin{equation} \label{double_roots}
    a^2_\infty = a_s^2 = \Gamma - 1. \hfill \mbox{(limit)}
\end{equation}
From (\ref{r_s}) we then have
\begin{equation}
    r_s = \frac{3 \Gamma - 2}{2 \Gamma - 2} M, \hfill \mbox{(limit)}
\end{equation}
which results in values outside the black-hole horizon at $r = 2 M$ for all $\Gamma < 2$.  

\begin{figure}
    \centering
    \includegraphics[width = 0.45 \textwidth]{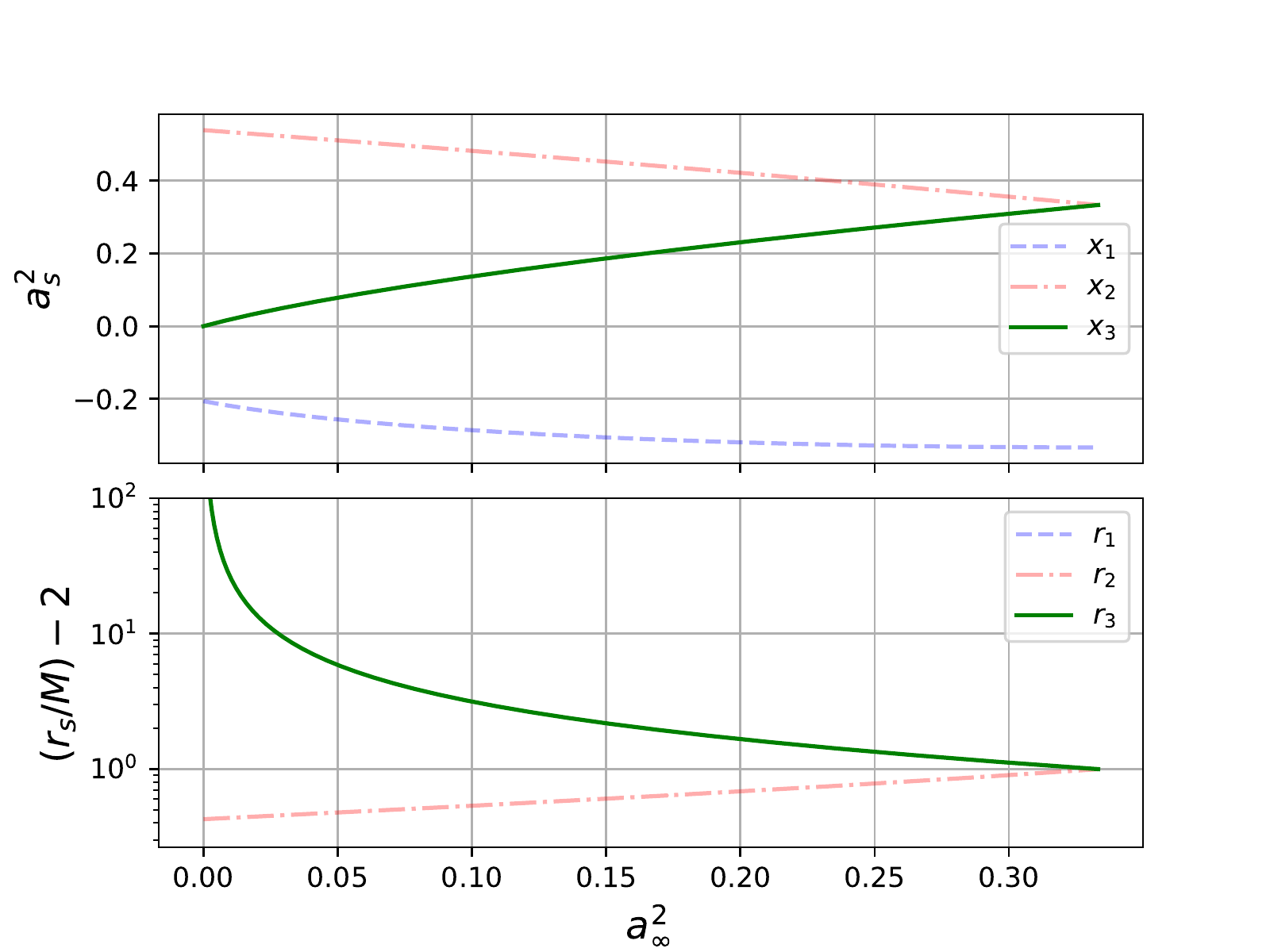}
    \caption{Upper panel: The three roots $x_1$, $x_2$, and $x_3$ to the cubic equation (\ref{gen_cubic}) as a function of $a_\infty^2$, for $\Gamma = 4/3$.
    Only solution $x_3$, high-lighted by the dark color, represents a physical accretion solution; solution $x_1$ is negative and hence unphysical, while solution $x_2$ represents an outgoing (i.e.~wind) solution.  Note the double-root at $a_\infty^2 = a_s^2 = \Gamma - 1 = 1/3$.  Lower panel: the critical radius $r_s$ associated with the roots shown in the upper panel.
    }
    \label{fig:cubic_solutions_4ov3}
\end{figure}

\begin{figure}
    \centering
    \includegraphics[width = 0.45 \textwidth]{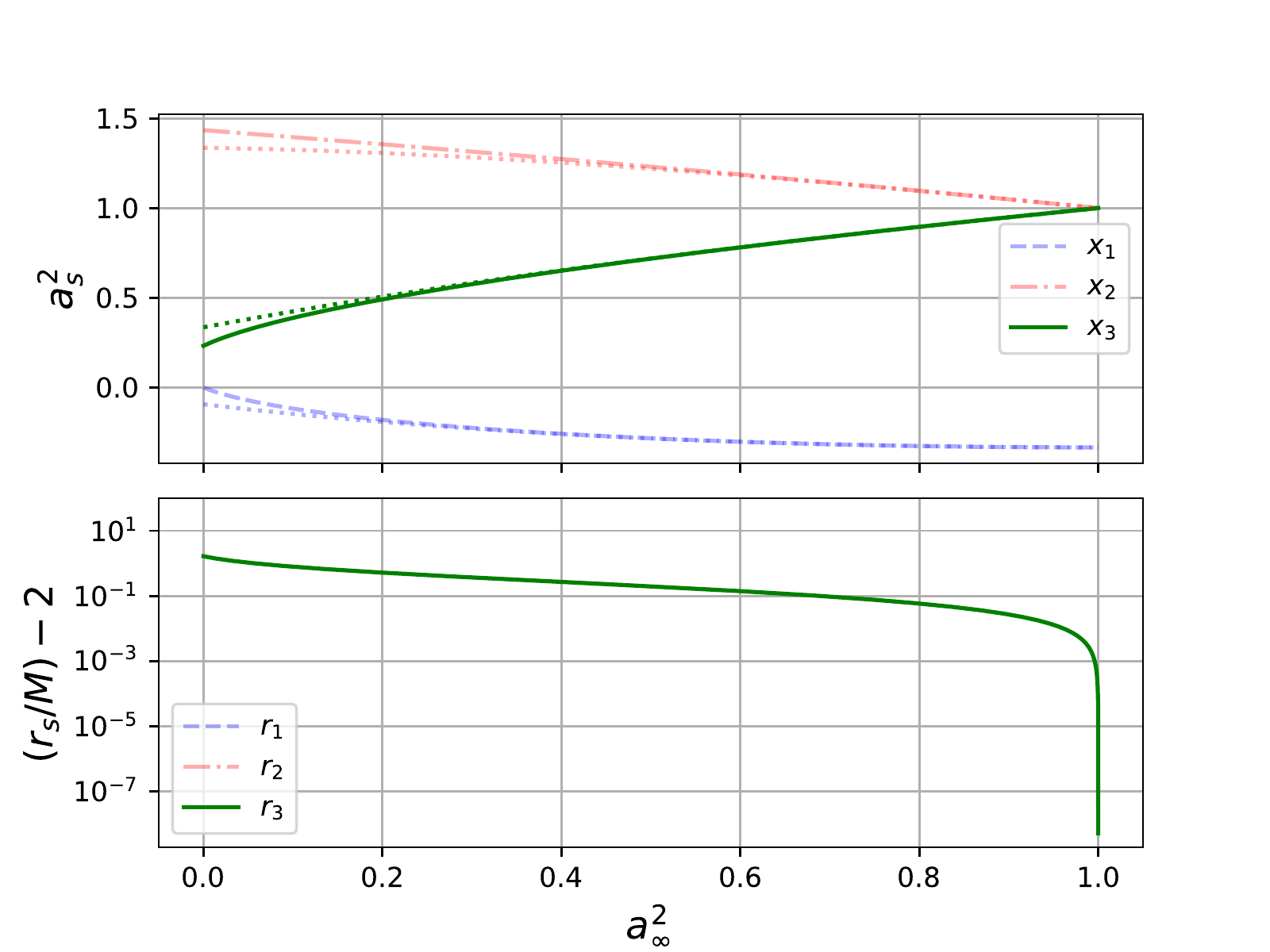}
    \caption{Same as Fig.~\ref{fig:cubic_solutions_4ov3}, but for $\Gamma = 2$.  The physical accretion solution is again given by $x_3$; note that this solution takes a nonzero value in the limit $a^2_\infty \rightarrow 0$ (see \ref{as_stiff_lowdensity}).  The dotted lines represent the high-density expansions (\ref{as_stiff_highdensity}).}  
    \label{fig:cubic_solutions}
\end{figure}

We now compute general solutions to the cubic equation (\ref{gen_cubic}) following Section 5.6 in \cite{NumRec}.  We start by defining
\begin{equation}\label{Q}
Q \equiv \frac{A^2 - 3B}{9} = \frac{(3 \Gamma - 2)^2}{81}
\end{equation}
and
\begin{equation} \label{R}
\begin{split}    
R & \equiv  \frac{2A^3-9AB+27C}{54} \\
 & =  \frac{1}{1458} \Big(54 \, \Gamma^3 - 351 \,\Gamma^2 + 558 \, \Gamma  
+  486 \, a_\infty^2 ( \Gamma - 1) - 243 \, a_\infty^2 - 259 \Big).
\end{split}
\end{equation}
and distinguish two different cases depending on whether the quantity
\begin{equation} \label{RminusQ}
\begin{split}
R^2 - Q^3 & = \frac{(1 + a_\infty^2 - \Gamma)^2}{8748} \Big(243 \, \big( a_\infty^4 - 2\, a_\infty^2 (\Gamma - 1) \big) \\
& ~~~~~- 
(5 - 3 \Gamma)^2 (12 \Gamma - 11) \Big)
\end{split}
\end{equation}
is positive or negative.  Solving for a root of (\ref{RminusQ}) we see that we have $R^2 - Q^3 < 0$ whenever
\begin{equation} \label{a_inf_condition}
a_\infty^2 < \Gamma - 1 + \frac{2 \sqrt{3}}{27} 
\left( 3 \Gamma - 2 \right)^{3/2}.
\end{equation}
Evidently, this condition holds for all $a_\infty^2 \leq \Gamma - 1$, so that $R^2 - Q^3$ is negative for all physically viable solutions.  The cubic equation (\ref{gen_cubic}) then has three roots that are given by what is sometimes called Cardano's formula,
\begin{equation}\label{cubic_roots}
    \begin{split}
       x_1 & = -2\sqrt{Q}\cos\left(\frac{\theta}{3}\right) - \frac{A}{3} \\
       x_2 & = -2\sqrt{Q}\cos\left(\frac{\theta+2\pi}{3}\right) - \frac{A}{3} \\
       x_3 & = -2\sqrt{Q}\cos\left(\frac{\theta-2\pi}{3}\right) - \frac{A}{3},
    \end{split}
\end{equation} 
where
\begin{equation}\label{theta}
    \theta = \arccos\left(\frac{R}{Q^{3/2}}\right).
\end{equation}
Since, in our case, all the coefficients in (\ref{gen_cubic}) are real, the three roots $x_1$, $x_2$, and $x_3$ are also real.  We show examples of these solutions, together with their associated critical radii (\ref{r_s}), for a representative soft EOS with $\Gamma = 4/3$ in Fig.~\ref{fig:cubic_solutions_4ov3} and for a representative stiff EOS with $\Gamma = 2$ in Fig.~\ref{fig:cubic_solutions}.  

For $\Gamma \leq 2$, physically viable solutions are restricted by \eq~(\ref{amax}), which guarantees causality everywhere, i.e.~$a < 1$, but for $\Gamma > 2$ we also need to impose causality explicitly in addition to (\ref{amax}).  We have found that this condition will hold only for sufficiently small values of $a_\infty$.   Since, for these solutions, we have also found $a_s > a_\infty$, as one might expect, we can compute an approximate upper limit on $a_\infty$ by inserting $a_s = 1$ into \eq~(\ref{gen_quadratic}).  This then yields
\begin{equation} \label{as_one}
    a_{\infty, {\rm max}}^2 = 3 - \Gamma.  \hfill (\Gamma > 2)
\end{equation}
Here we have picked the ``-" solution in (\ref{gen_quadratic}), since the ``+" solution, $3 \Gamma - 5$, is not relevant in this regime.  We also observe that $a_s$ increases monotonically with $a_\infty$ for viable accretion solutions (see, e.g., Figs.~\ref{fig:profile4o3} and \ref{fig:profile2}), hence $a_\infty$ will be even smaller than (\ref{as_one}) for $a_s < 1$.  Condition (\ref{as_one}) is not sufficient, however, to guarantee that $a < 1$ for all values of $r$, so the true upper limit on $a_\infty$ may be smaller.  We therefore conclude from (\ref{as_one}) that physically viable solutions do not exist at least for $\Gamma \geq 3$.  


We observe in Figs.~\ref{fig:cubic_solutions_4ov3} and \ref{fig:cubic_solutions} that,
for $\Gamma = 4/3$, $a_s^2 \rightarrow 0$ as $a^2_\infty \rightarrow 0$, while, for $\Gamma = 2$, $a^2_s$ approaches a finite, nonzero value in this limit.  Similarly, $r_s$ grows without bound as $a^2_\infty \rightarrow 0$ for $\Gamma = 4/3$, but approaches a finite value for $\Gamma = 2$.

\begin{figure}
    \centering
    \includegraphics[width = 0.45 \textwidth]{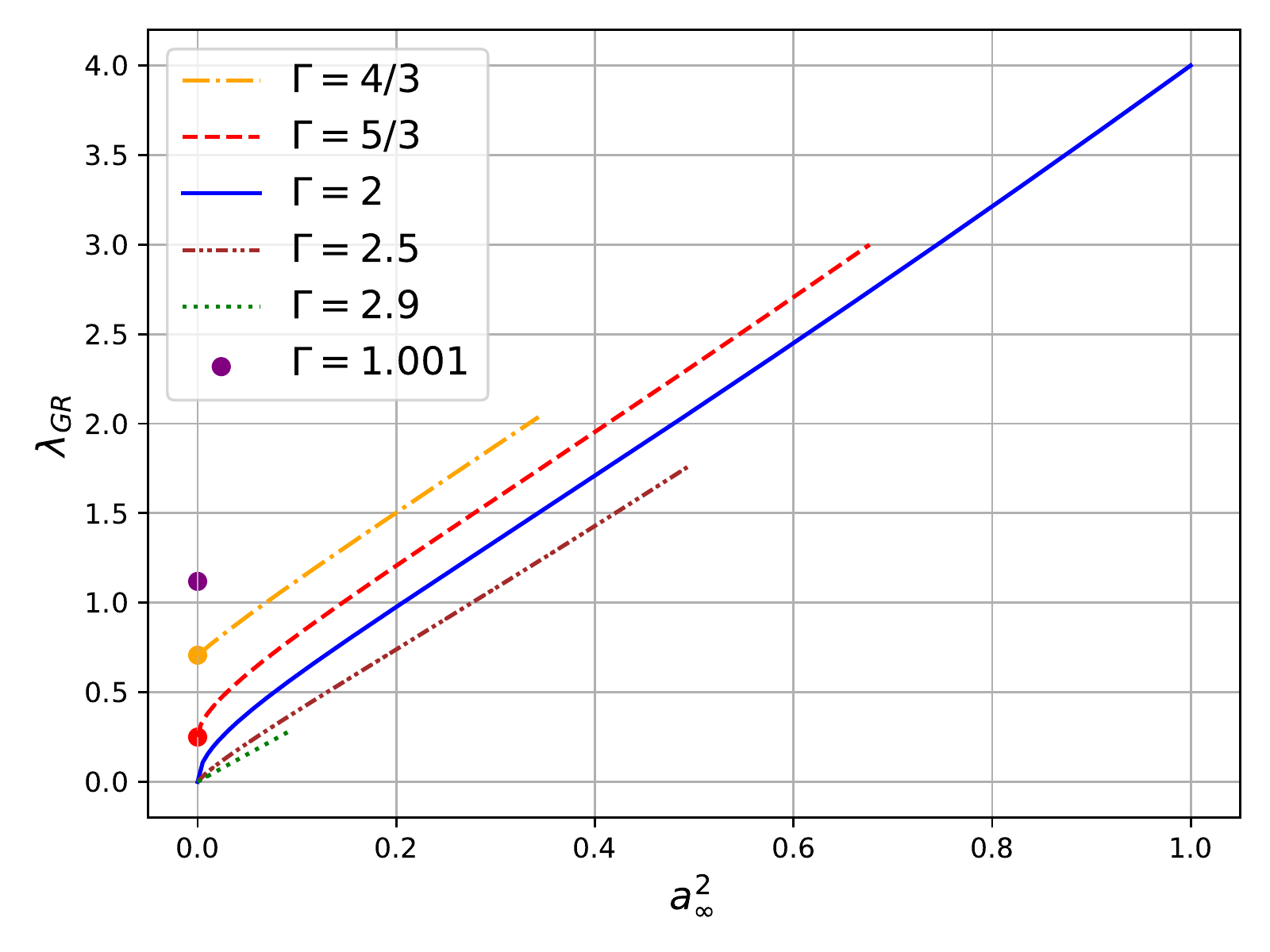}
    \caption{Values of the accretion rate eigenvalues $\lambda_{\rm GR}$ for different values of the adiabatic index $\Gamma$, as a function of $a_\infty^2$.  In the Newtonian limit $a^2_\infty \ll 1$ the relativistic eigenvalues approach their Newtonian counterparts for $1 \leq \Gamma \leq 5/3$, shown as the dots for $a^2_\infty \rightarrow 0$.  For $\Gamma > 5/3$, the accretion rate eigenvalues drop to zero in this limit.}
    \label{fig:lambdas}
\end{figure}

For a given value of $\Gamma$ and $a_\infty$ we can now insert $a_s$ into (\ref{lambda_GR}) to obtain the relativistic accretion eigenvalue $\lambda_{\rm GR}$.  We show examples of $\lambda_{\rm GR}$ as a function of $a_\infty^2$, for different values of $\Gamma$, in Fig.~\ref{fig:lambdas}.  For $\Gamma \leq 2$, these solutions extend up to $a_\infty^2 = a_{\rm max}^2 = \Gamma - 1$, while for $\Gamma \geq 2$ they are limited by the existence of solutions with $a_s < 1$, i.e.~condition (\ref{as_one}).  For $1 \leq \Gamma \leq 5/3$ we also include their Newtonian counterparts $\lambda_s$ as the dots for $a_\infty^2 \rightarrow 0$ (see \ref{lambda_s}, or Table 14.1 in ST).  As expected, the relativistic values approach the Newtonian ones in this Newtonian limit.  For $\Gamma > 5/3$ on the other hand, we observe that the values of $\lambda_{\rm GR}$ drop to zero as $a^2_\infty \rightarrow 0$.

\begin{figure}
    \centering
    \includegraphics[width = 0.45 \textwidth]{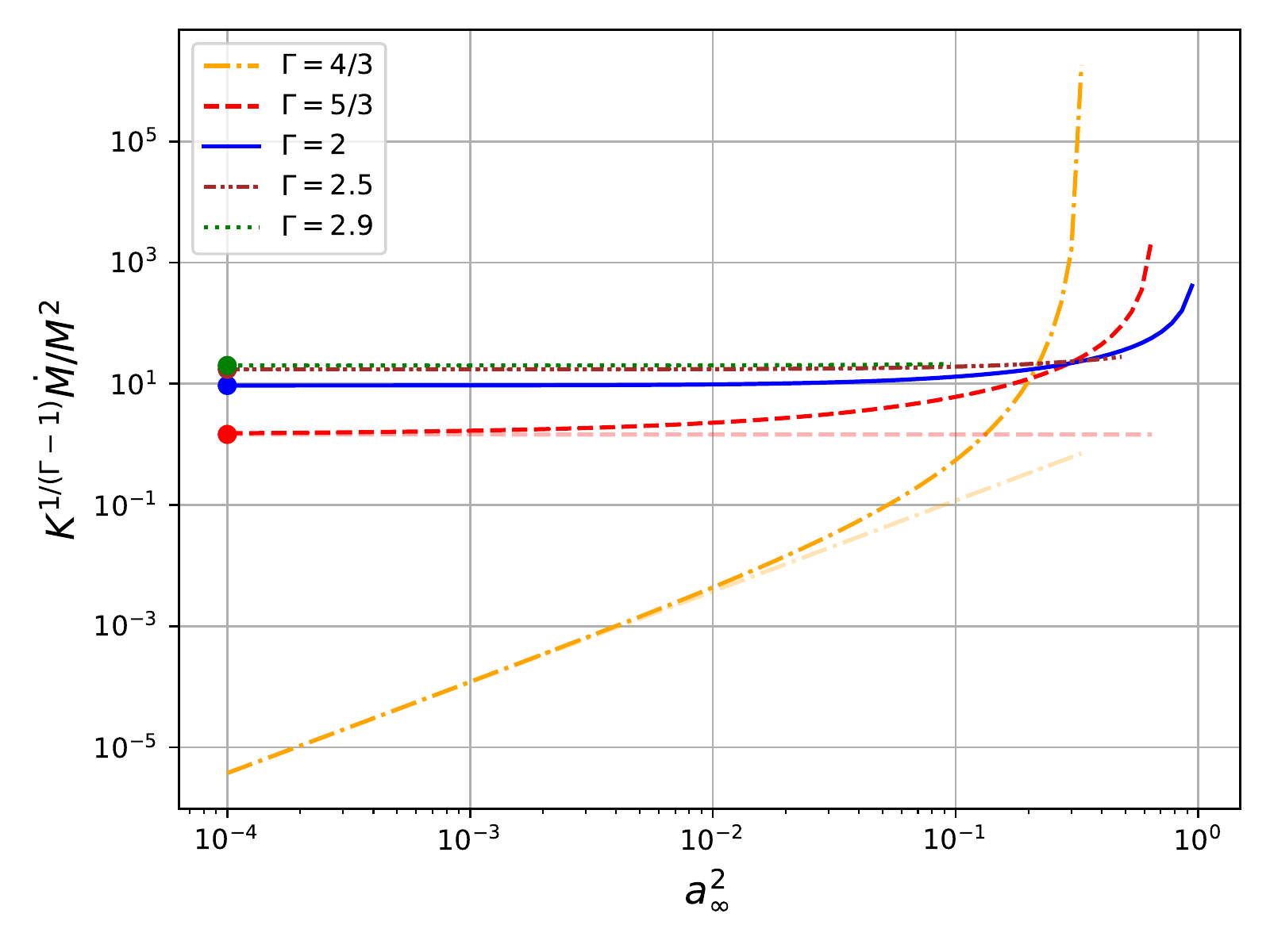}
    \caption{Accretion rates $K^{1/(\Gamma - 1)} \dot M/ M^2$ as a function of $a_\infty^2$ (with $10^{-4} \leq a_\infty^2 < a_{\rm max}^2 = \min\,(\Gamma - 1, 3 - \Gamma)$) for different values of $\Gamma$.  For $\Gamma < 5/3$ these accretion rates approach zero as $a^2_\infty \rightarrow 0$, but for $\Gamma \geq 5/3$ they approach a finite, non-zero value.  For the former we include the low-density power-law fits (\ref{mdot_soft_smalldensity}) as the faint line, while for the latter we include the minimum accretion rates (\ref{Mdot_min}) as the dots.
    }
    \label{fig:accretion}
\end{figure}

Given values of $\lambda_{\rm GR}$, and expressing $\rho_{0\infty}$ in terms of $a^2_\infty$ using (\ref{a2_of_rho0}), we can then find accretion rates $\dot M$ from (\ref{Mdot}).  We show results for different values of the adiabatic exponent $\Gamma$ in Fig.~\ref{fig:accretion}.  For $\Gamma < 5/3$, the accretion rates drop to zero as $a^2_\infty \rightarrow 0$, but for $\Gamma \geq 5/3$ they approach a finite, non-zero value in this limit (as long as $K$ is finite; see Section \ref{sec:small_a_large_gamma} for examples for which this condition does or does not apply).

\begin{figure}
    \centering
    \includegraphics[width = 0.45 \textwidth]{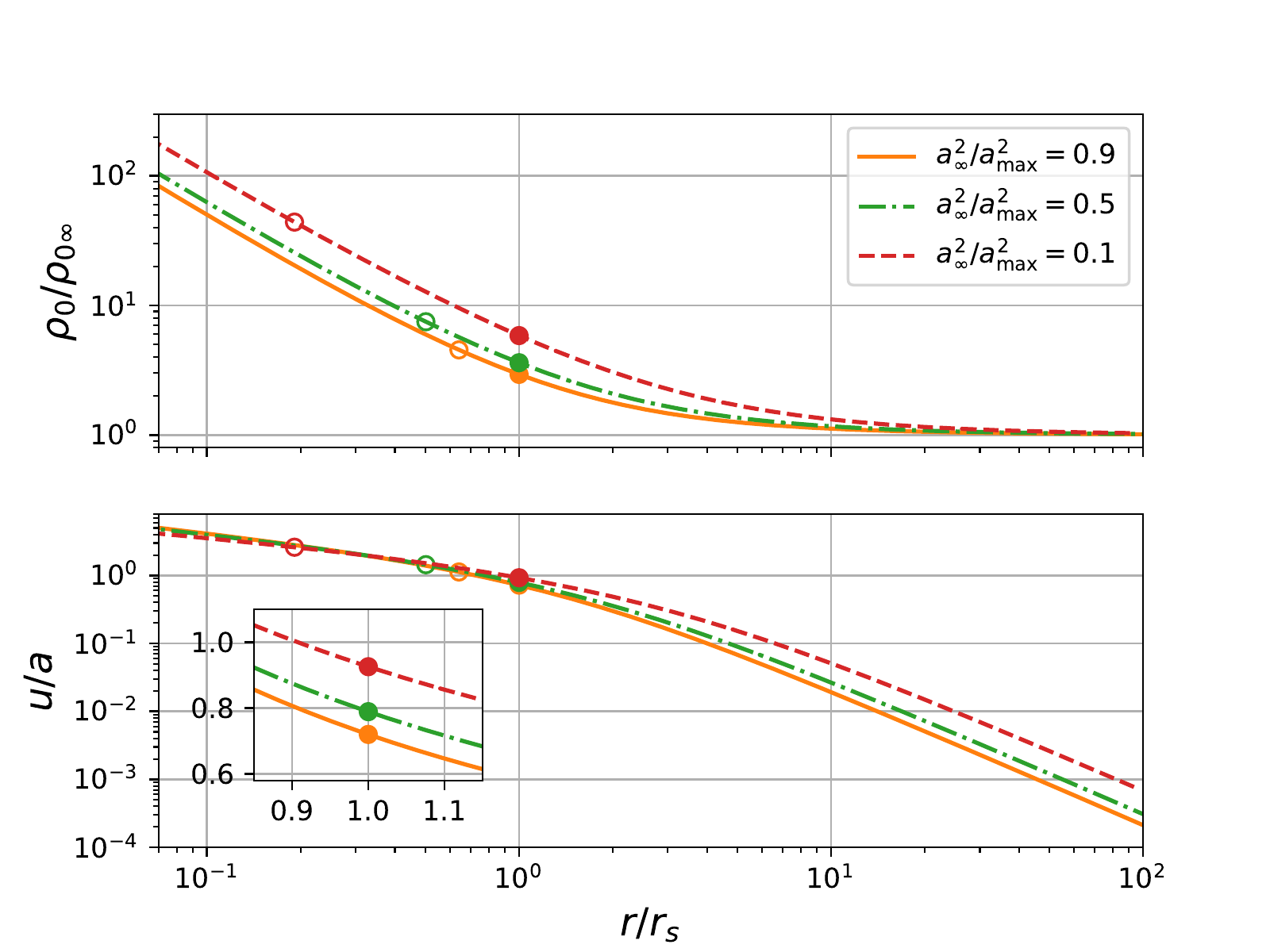}
    \caption{Profiles of the rest-mass density $\rho_0/\rho_{0\infty}$ (upper panel) and the ratio between the fluid velocity velocity $u$ and sound speed $a$ (lower panel) for $\Gamma = 4/3$, and for three representative values of the asymptotic sound speed $a_{\infty}$.  Here $a^2_{\rm max} = \Gamma - 1 = 1/3$ (see \ref{amax}).  The solid dots denote values at the critical radius $r_s$, while the open circles mark the location of the event horizon at $r = 2M$.
    }
    \label{fig:profile4o3_large_a}
\end{figure}

\begin{figure}
    \centering
    \includegraphics[width = 0.45 \textwidth]{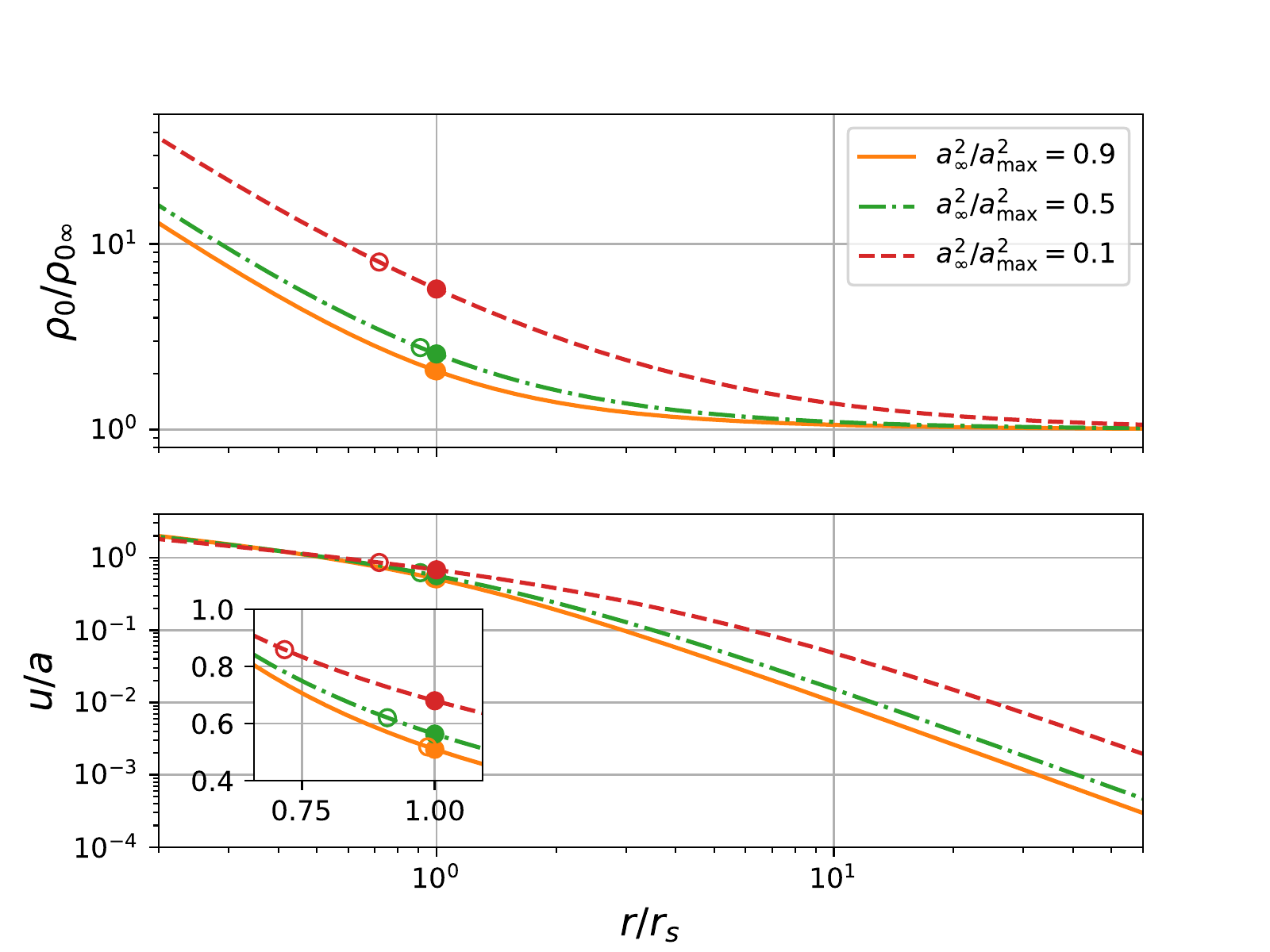}
    \caption{Same as Fig.~\ref{fig:profile4o3_large_a}, but for $\Gamma = 2$.
    }
    \label{fig:profile2_large_a}
\end{figure}

\begin{figure}
    \centering
    \includegraphics[width = 0.45 \textwidth]{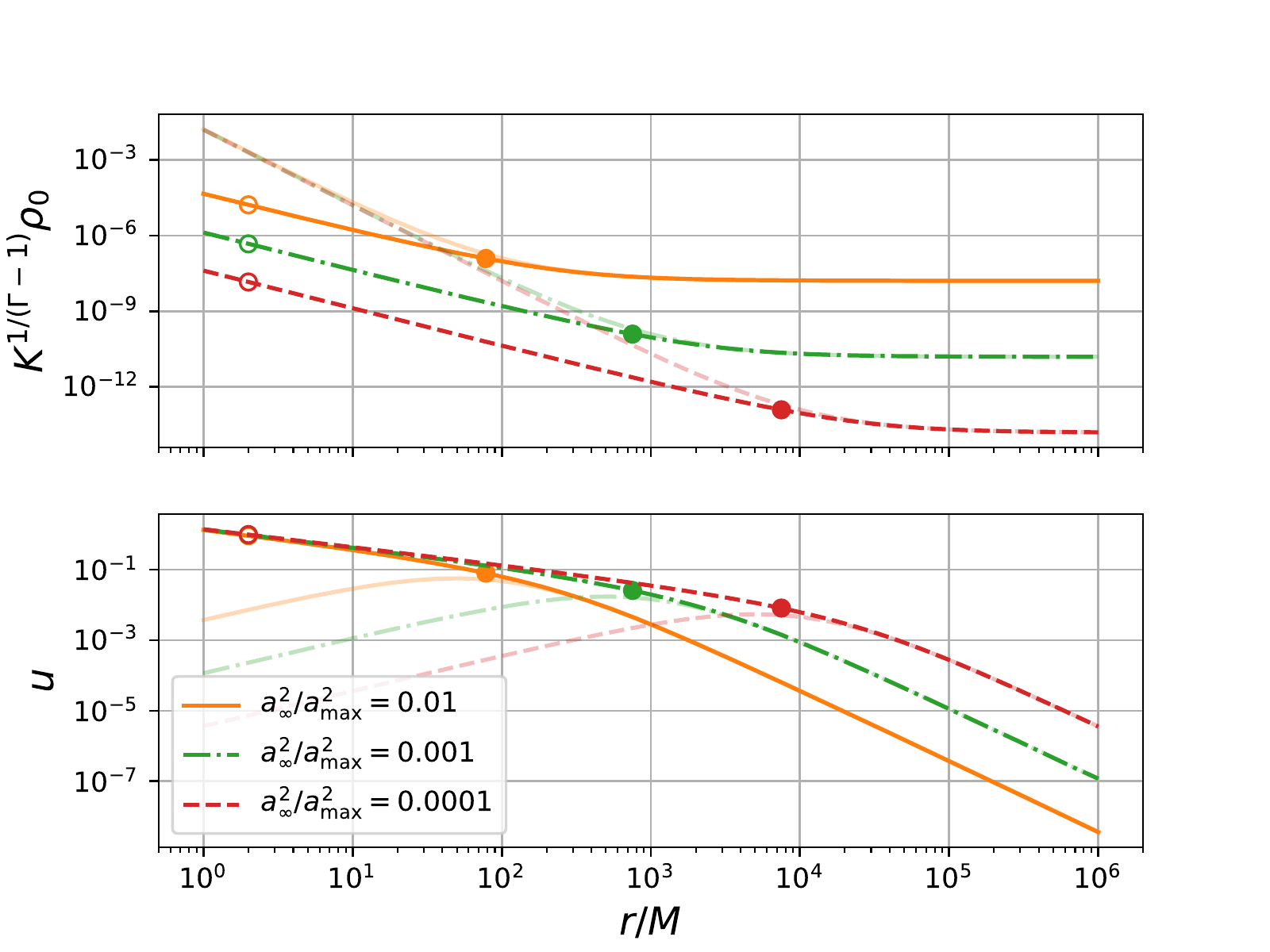}
    \caption{Same as Fig.~\ref{fig:profile4o3_large_a}, except for small values of the asymptotic sound speed.  We also plot $K^{1/(\Gamma - 1)} \rho_0$ and $u$ here, rather than $\rho_0 / \rho_{0\infty}$ and $u/a$, and rescale the radius $r$ with regard to the black hole mass $M$ rather than the critical radius $r_s$. The horizon, marked by open circles, is at $r/M = 2$, while the solid dots denote values at the critical radius $r_s$. The faint lines, which mark the approximate asymptotic solutions (\ref{asymptotic_profile}), overlap the actual solutions at large radii.  Note that the density decreases everywhere as $a_\infty^2 \rightarrow 0$.}  
    \label{fig:profile4o3}
\end{figure}

\begin{figure}
    \centering
    \includegraphics[width = 0.45 \textwidth]{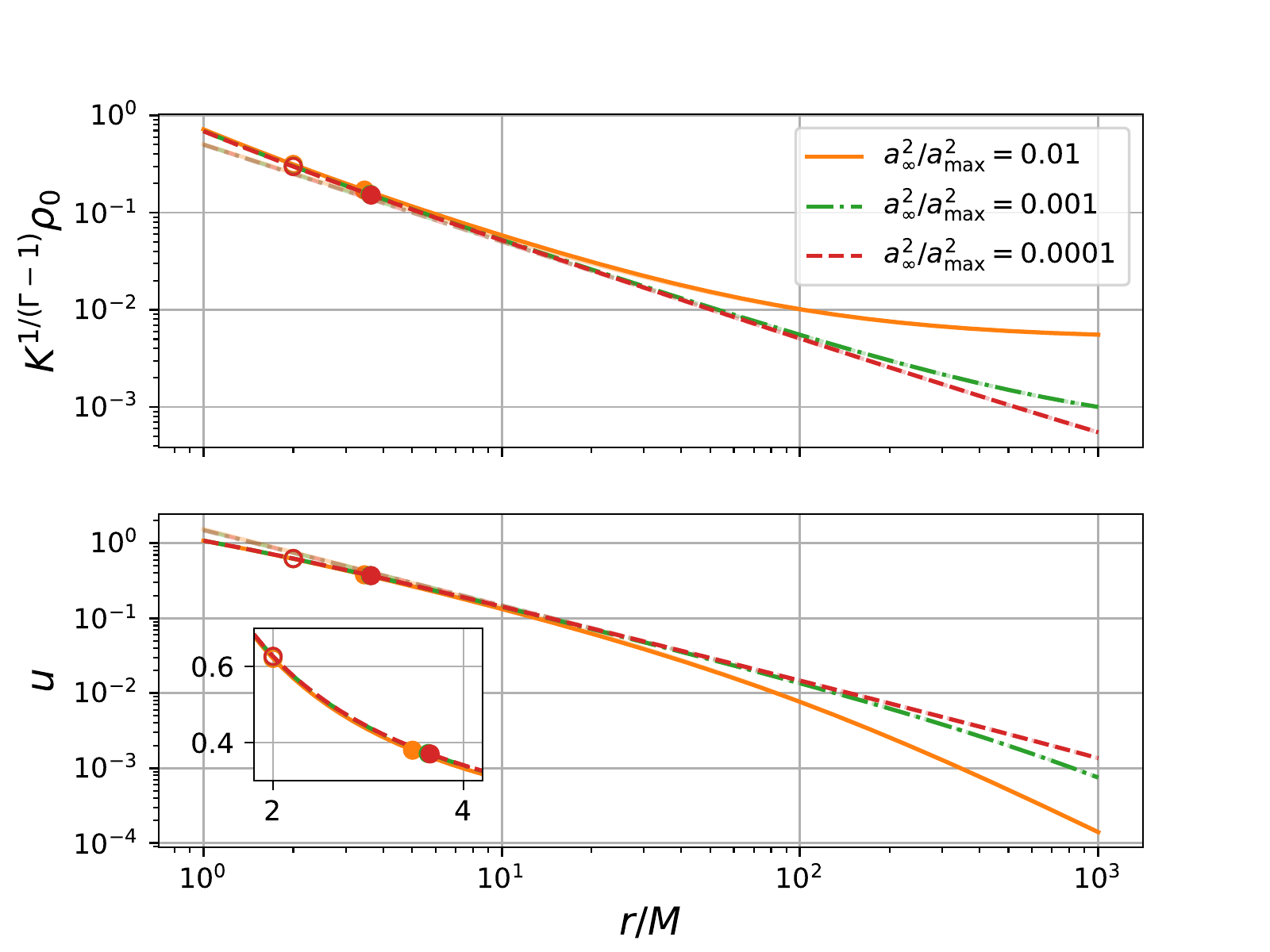}
    \caption{Same as Fig.~\ref{fig:profile4o3}, but for $\Gamma = 2$.  Note that the critical radius, marked by the solid dots, approaches a fix point, and that the density profile approaches the asymptotic power-law (\ref{powerlaw_profile}), as $a_{\infty}^2 \rightarrow 0$.  The approximate asymptotic solutions (\ref{asymptotic_profile}) are included as faint lines, but are difficult to see since they agree quite well with the actual solution, down to small radii $r$. 
    }
    \label{fig:profile2}
\end{figure}

Finally, it is useful to consider profiles of the rest-mass density $\rho_0$ and fluid velocity $u$.  This involves integrating the relativistic Euler equation and continuity equation (\ref{prime}), starting with values of the fluid variables at the critical radius (see Appendix \ref{sec:profiles:equations}).  We show results for $\Gamma = 4/3$ and $\Gamma = 2$ in Figs.~\ref{fig:profile4o3_large_a} through \ref{fig:profile2}.  Figs.~\ref{fig:profile4o3_large_a} and
\ref{fig:profile2_large_a} show results for values of $a_\infty^2 / a_{\rm max}^2$ between 0.1 and 0.9 (where $a_{\rm max}^2 = \Gamma - 1$ is given by \ref{amax}), while in Figs.~\ref{fig:profile4o3} and \ref{fig:profile2} we focus on small asymptotic sound speeds $a_\infty^2 / a_{\rm max}^2$ between 0.01 and 0.0001.

As before, we notice qualitative differences between the behavior for soft and stiff EOSs.  Specifically, as we reduce the asymptotic density $\rho_{0\infty}$, the density decreases {\em everywhere} for $\Gamma = 4/3$, but approaches the asymptotic power-law (\ref{powerlaw_profile}) for $\Gamma = 2$.  Simultaneously, the critical radius $r_s$ keeps increasing as $\rho_{0\infty} \rightarrow 0$ for $\Gamma = 4/3$, and instead approaches a fixed point for $\Gamma = 2$.  All of this is consistent with our previous observations about the critical point and accretion rate.

\section{Limiting cases}

Given the qualitative differences that we observe for soft and stiff EOS it is instructive to examine this behavior more carefully by considering limiting cases.

\subsection{Low asymptotic sound speeds}

We start with the limit of low asymptotic sound speeds $a^2_\infty~\ll~1$ \citep[see][]{Beg78}. In this limit, the rest-mass density can be approximated from (\ref{a2_of_rho0}),
\begin{equation} \label{density_small_a}
    \rho_0 = \left( \frac{a^2}{\Gamma K} \right)^{1/(\Gamma - 1)} \left( 1 + \mathcal{O}(a^2) \right).  
    \hfill (a^2 \ll 1)
\end{equation}
and we may express the accretion rate (\ref{Mdot}) as
\begin{equation} \label{mdot_small_a}
    \dot M = 4 \pi \lambda_{GR} \, \frac{M^2}{(\Gamma K)^{1/(\Gamma - 1)}} \, a_\infty^{(5 - 3\Gamma)/(\Gamma - 1)} \left(1 + \mathcal{O}(a_\infty^2)\right).
    \hfill (a^2_\infty \ll 1)
\end{equation}

We first note that, for $a_\infty = 0$, the coefficient $C$ in (\ref{cubic_components}) vanishes, so that the solutions to the cubic equation (\ref{gen_cubic}) are given by
\begin{equation}
    \begin{split}
        \bar x_I & = 0 \\
        \bar x_{II} & = \Gamma - 7/6 + \left(12 \Gamma - 11 \right)^{1/2} / 6 \\ 
        \bar x_{III} & = \Gamma - 7/6 - \left(12 \Gamma - 11 \right)^{1/2} / 6.  
    \end{split}
    \hfill (a_\infty^2 = 0)
\end{equation}
Here we have used Roman subscripts, because which one of these three roots should be identified with which of the three roots in (\ref{cubic_roots}) depends on the value of $\Gamma$ (see below).  

To leading order in $a_\infty^2$, the roots to the cubic equation can now be written as
\begin{equation} \label{a_small_expansion}
    \begin{split}
        x_I & = a_I \,a_\infty^2 \\
        x_{II} & = \bar x_{II} + a_{II}\, a_\infty^2 \\
        x_{III} & = \bar x_{III} + a_{III} \, a_\infty^2. \\
    \end{split}
    \hfill (a_\infty^2 \ll 1)
\end{equation}
where the coefficients $a_I$, $a_{II}$, and $a_{III}$ are given by
\begin{equation} \label{a_small_coef}
    \begin{split}
        a_I & = \frac{2}{5 - 3 \Gamma} \\
        a_{II} & = - \frac{2}{5 - 3 \Gamma} \, 
        \frac{\bar x_{II} + A}{\bar x_{II} - \bar x_{III}} \\
        a_{III} & = \frac{2}{5 - 3 \Gamma} \, 
        \frac{\bar x_{III} + A}{\bar x_{II} - \bar x_{III}}.
    \end{split}
\end{equation}
We now observe that $B$, $\bar x_{III}$, the coefficients (\ref{a_small_coef}), as well as the exponent of the first term in (\ref{lambda_GR}) all change sign at $\Gamma = 5/3$.  The coefficients (\ref{a_small_coef}) diverge for $\Gamma = 5/3$, meaning that the expansion does not converge in this case, and that we have to treat the cases $\Gamma \leq 5/3$ and $\Gamma > 5/3$ separately.

\subsubsection{Soft equations of state: $\Gamma \leq 5/3$}

For soft EOSs with $\Gamma < 5/3$, the exponent of the first term in the accretion eigenvalue (\ref{lambda_GR}) is positive.  For $\lambda_{GR}$ to remain finite in the limit $a^2_\infty \rightarrow 0$, we therefore need to choose $a_s^2 = x_I$.  From Fig.~\ref{fig:cubic_solutions_4ov3} we see that we can identify this solution with $x_3$ in (\ref{cubic_roots}).  We then have
\begin{equation} \label{as_soft}
\frac{a_s}{a_\infty} = a_I^{1/2} + {\mathcal O}(a_\infty^2) = \left(\frac{2}{5 - 3 \Gamma} \right)^{1/2} + {\mathcal O}(a_\infty^2),
\hfill (\Gamma < 5/3)
\end{equation}
in accordance with \eq~(\ref{as_o_ainfty}) above.  For $\Gamma = 5/3$ we find, from direct expansion of (\ref{setup_cubic}),
\begin{equation} \label{a_s_5o3}
a_s^2 = \frac{2}{3} a_\infty \hfill (a_\infty \ll 1,~\Gamma = 5/3) 
\end{equation}
(see ST, Ex.~G.1).  Substituting \eq~(\ref{as_soft}) into \eq~(\ref{r_s}) yields, for $\Gamma < 5/3$,
\begin{equation}
    r_s = \frac{5 - 3 \Gamma}{4} \, \frac{M}{a_\infty^2}, 
    \hfill (a_\infty \ll 1,~\Gamma < 5/3)
\end{equation}
which agrees with the Newtonian result, while for $\Gamma = 5/3$ we insert (\ref{a_s_5o3}) into (\ref{r_s}) to obtain
\begin{equation} \label{r_s_5o3}
    r_s = \frac{3}{4} \, \frac{M}{a_\infty}. 
    \hfill (a_\infty \ll 1,~\Gamma = 5/3)
\end{equation}
Note that $r_s$ in (\ref{r_s_5o3}) is greater than zero, in contrast to the Newtonian value (see ST, Ex.~G.1). 

We now insert (\ref{as_soft}) into (\ref{lambda_GR}) and recover, not surprisingly, the Newtonian result (\ref{lambda_s}),
\begin{equation} \label{lambda_limit_soft}
    \lambda_{GR} = \frac{1}{4} a_I^{(5-3\Gamma)/(2\Gamma - 2)} + \mathcal{O}(a_\infty^2)
    = \lambda_s + \mathcal{O}(a_\infty^2).
    \hfill (\Gamma \leq 5/3)
\end{equation}
Note, in particular, that $\lambda_{GR}$ approaches a finite, non-zero value as $a^2_\infty \rightarrow 0$ (as shown in Fig.~\ref{fig:lambdas}, where the above values are included as dots).  Also, even though the expansion (\ref{a_small_expansion}) does not converge for $\Gamma = 5/3$, $\lambda_s$ does take the finite value of 1/4 in this limiting case, as \eq~(\ref{lambda_s}) reveals.

We can now evaluate the accretion rate (\ref{mdot_small_a}) to find, to leading order,
\begin{equation} \label{mdot_soft_smalldensity}
    \dot M 
    = \frac{\pi M^2}{(\Gamma K)^{1/(\Gamma -1)}} 
    \left( \frac{2 a^2_\infty}{5 - 3 \Gamma} \right)^{(5 - 3 \Gamma)/(2\Gamma - 2)}.
\hfill (a_\infty^2 \ll 1,~\Gamma \leq 5/3)
\end{equation}
For $\Gamma < 5/3$ this expression predicts that the accretion rate will vanish in the limit $a^2_\infty \rightarrow 0$.  This is shown in Fig.~\ref{fig:accretion}, where we have included the leading-order result (\ref{mdot_soft_smalldensity}) as the faint lines.  For $\Gamma = 5/3$, however, the accretion rate (\ref{mdot_soft_smalldensity}) approaches a non-zero value as $a^2_\infty \rightarrow 0$ -- similar to the behavior that we will encounter for stiff EOSs in the following Section.

\subsubsection{Stiff equations of state: $\Gamma > 5/3$}
\label{sec:small_a_large_gamma}

Now consider stiffer EOSs with $\Gamma > 5/3$.  Since $a_I < 0$ in this case, $x_I$ in (\ref{a_small_expansion}) no longer represents a physically viable solution.  Also, since $x_{II} > 1$ for all $\Gamma > 5/3$, we now choose $a_s^2 = x_{III}$, i.e.  
\begin{equation} \label{as_stiff_lowdensity}
    a_s^2 = \bar x_{III} + a_{III} \, a^2_\infty + \mathcal{O}(a_\infty^4).
\end{equation}
From Fig.~\ref{fig:cubic_solutions} we see that we can again identify this solution with $x_3$ in (\ref{cubic_roots}).  Note that $a_s$ takes a non-zero limiting value of $\bar x_{III}^{1/2}$ even in the limit $a_\infty \rightarrow 0$ (see Fig.~\ref{fig:cubic_solutions} for an example).  This indicates that accretion flow profiles with smaller values of $a_s$ cannot be extended to infinity, meaning that such solutions are local rather than global.  Whether or not solutions allow global extensions was studied in detail by \cite{ChaMS16}; for our limiting value of $a_s = \bar x_{III}^{1/2}$ their variable $L^2$ takes the value of unity, which they demonstrate is the limiting value allowing a global accretion flow extending from the horizon to the asymptotic region.  Note also that $\bar x_{III} \geq 1$ for $\Gamma \geq 3$, meaning that no physical solutions exist in this regime.

Inserting (\ref{as_stiff_lowdensity}) into (\ref{r_s}) yields
\begin{align}
    r_s & = \frac{1 + 3 \bar x_{III}}{2 \bar x_{III}} \, M \nonumber \\ 
    & = \frac{3 \Gamma - 5/2 - (12 \Gamma - 11)^{1/2}/2}{2 \Gamma - 7/3 - (12 \Gamma - 11)^{1/2}/3} \, M, ~~~~~~~ (a_\infty^2 \ll 1,~\Gamma > 5/3) 
\end{align}
which, to leading order, is independent of $a_\infty^2$.  Also, since $a_\infty / a_s = \bar x_{III}^{-1/2} a_\infty  + \mathcal{O}(a_\infty^3)$, the accretion eigenvalues (\ref{lambda_GR}) become
\begin{equation} \label{lambda_limit_stiff}
    \lambda_{GR} = \bar \lambda \, a_\infty^{(3\Gamma - 5)/(\Gamma - 1)} 
        \left( 1 + \mathcal{O}(a_\infty^2) \right) 
    \hfill (\Gamma > 5/3)
\end{equation}
with 
\begin{equation} \label{lambda_bar}
\bar \lambda = \bar x_{III}^{(5 - 3\Gamma)/(2 (\Gamma - 1))}
    \left( \frac{\Gamma - 1}{\Gamma - 1 - \bar x_{III}}\right)^{1/(\Gamma - 1)} 
    \frac{(1 + 3 \bar x_{III})^{3/2}}{4}. 
\end{equation}
Unlike their counterparts for $\Gamma \leq 5/3$, the accretion eigenvalues (\ref{lambda_limit_stiff}) vanish in the limit of $a^2_\infty \rightarrow 0$.  In fact, combining the results (\ref{lambda_limit_soft}) and (\ref{lambda_limit_stiff}) we see that, for $a^2_\infty \rightarrow 0$, the accretion eigenvalue drops {\em discontinuously} from its (non-vanishing) Newtonian values $\lambda_s$ for $\Gamma \leq 5/3$ to zero for $\Gamma > 5/3$,
\begin{equation} \label{lambda_limit}
\lambda_{GR} =
\left\{ \begin{array}{ll}
\lambda_s~~~~~ & \Gamma \leq 5/3 \\
0 & \Gamma > 5/3
\end{array}
\right.
\hfill (a_\infty^2 \ll 1)
\end{equation}
(see also Fig.~\ref{fig:lambdas}).  While this behavior may seem surprising, it is, in fact, necessary in order to keep the accretion rate (\ref{mdot_small_a}) finite for $\Gamma > 5/3$.  Inserting (\ref{lambda_limit_stiff}) into (\ref{mdot_small_a}) we see that the leading-order dependence on $a_\infty$ now cancels out, and we obtain, to leading order,
\begin{equation} \label{Mdot_min}
\dot M = 4 \pi \bar \lambda \, \frac{M^2}{(\Gamma K)^{1/(\Gamma - 1)}}. 
    \hfill (a_\infty^2 \ll 1,~\Gamma \geq 5/3)
\end{equation}
In the limit of $\Gamma \rightarrow 5/3$, \eqs~(\ref{Mdot_min}) and (\ref{mdot_soft_smalldensity}) yield the same result.

Remarkably, \eq~(\ref{Mdot_min}) indicates that, for stiff equations of state with $\Gamma > 5/3$, the accretion rate approaches a non-zero limiting value as the sound speed and density at infinity approach zero (see Fig.~\ref{fig:accretion}, where we have included the limiting accretion rates (\ref{Mdot_min}) as dots for $\Gamma \geq 5/3$).  While we do not provide a formal proof here, it can be seen in Fig.~\ref{fig:accretion} that these limiting values represent minima.
We therefore conclude that, for stiff equations of state with $5/3 \leq \Gamma < 3$, there exists a {\em minimum} steady-state accretion rate, given by (\ref{Mdot_min}).  As we discuss in Appendix \ref{sec:profiles:asymptotics}, this minimum accretion rate is associated with a limiting density profile that, asymptotically, approaches the power-law (\ref{powerlaw_profile}). 

The minimum accretion rate given by (\ref{Mdot_min}) depends on the values of $M$, $\Gamma$, and $K$, where the latter two quantities are determined by the adopted EOS.  To clarify the novelty and applicability of this minimum rate, let us consider two cases involving the same $M$ and $\Gamma = 5/3$, but with different values of $K$ corresponding to two different EOSs. For an ideal Maxwell-Boltzmann gas, we have $P = \rho_0 k T/m$, where $T$ is the temperature, $k$ is Boltzmann's constant and $m$ is the mean mass of the gas particles. For $a^2_{\infty} \ll 1$ we have for the sound speed $a_{\infty}^2 = (5/3) \, kT_{\infty}/m$, which we shall set to be some fixed, finite, nonzero asymptotic value. According to (\ref{K}), as we consider adiabatic gases with successively lower asymptotic densities but with the same nonzero $a^2_{\infty}$, then as we let $\rho_{0 \infty} \rightarrow 0$ we have $K \approx a_{\infty}^2/(\Gamma \rho_{0 \infty}^{\Gamma-1}) \rightarrow \infty$.
According to (\ref{Mdot_min}) we then find that the minimum accretion rate also tends to zero, i.e, $\dot{M}_{\rm min} \rightarrow 0$. The vanishing of the accretion rate for vanishingly small asymptotic densities found here is not surprising; it is already evident from (\ref{Mdot}) in this case, where $a_{\infty}$ remains finite while $\rho_{0 \infty}$ falls to zero. 

But now consider an ideal, nonrelativistic, cold, degenerate gas for which $K$ is a positive constant determined by fundamental atomic constants (see, e.g., ST 2.3.22 for a degenerate electron gas and ST 2.3.27 for a degenerate neutron gas). Now according to (\ref{Mdot_min}) we have $\dot{M}_{\rm min} \rightarrow  constant > 0$  as  $\rho_{0 \infty} \rightarrow 0$. Thus $\dot{M}$ converges to $\dot{M}_{\rm min}$ independently of $\rho_{0 \infty}$ as $\rho_{0 \infty} \rightarrow 0$ and thereby tends to a universal, nonzero value that is given by fundamental constants. This result is rather surprising. In this case (\ref{Mdot}) also yields the correct answer: while now $a_{\infty}^2  \approx \Gamma K \rho_{0 \infty}^{\Gamma -1}$ tends to zero whenever $\rho_{0\infty}$ tends to zero, when $\Gamma = 5/3$ the densities appearing in the numerator and denominator cancel. The result is again the minimum accretion rate given by (\ref{Mdot_min}).  When the matter pressure is dominated by degenerate electrons, this mininum accretion rate is given by 
\begin{equation}
\begin{split}
    \dot M_{\rm min} & = \frac{3^{1/2}}{\pi} \left[ \frac{G^2 m_e^{3/2} m_u^{5/2} \mu_e^{5/2}}{\hbar^3} \right] M^2 \\
    & \simeq 0.41 \,\mu_e^{5/2} \, \frac{M_\odot}{\mbox{s}} \left( \frac{\mns}{M_\odot} \right)^2, 
    \end{split}
    \hfill (\mbox{degenerate electrons})
\end{equation}
where $m_e$ is the electron mass, $m_u$ the atomic mass unit, and $\mu_e$ the mean molecular weight per electron, while for degenerate neutrons it is
\begin{equation}
\begin{split}
    \dot M_{\rm min} & = \frac{3^{1/2}}{\pi} \left[ \frac{G^2 m_n^4}{\hbar^3}  \right] M^2 \\
    & \simeq 3.2 \times 10^4 \frac{M_\odot}{\mbox{s}} \left( \frac{\mns}{M_\odot} \right)^2,
    \end{split}
    \hfill (\mbox{degenerate neutrons})
\end{equation}
where $m_n$ is the neutron mass.  Note, of course, that we are ignoring effects of radiation, which might otherwise reduce the accretion rate below the above limits.

As another example, consider an EOS with $\Gamma = 2$, which is often employed in simple models of neutron stars and simulations of binary neutron star mergers to represent a cold, stiff nuclear EOS.  In this case, we have 
\begin{equation}
    \bar \lambda \simeq 1.49 \hfill (\Gamma = 2)
\end{equation}
and 
\begin{equation}
    \dot M_{\rm min} \simeq 9.39 \, \frac{M^2}{K}. \hfill (\Gamma = 2)
\end{equation}

\subsection{High asymptotic sound speeds}

We next consider the limit of large sound speeds, recognizing that fluids typically become stiff at high densities.  Deep in the core of a massive neutron star, for example, the sound speed may be close to the speed of light, i.e.~$a^2_\infty \lesssim 1$.  Specializing to $\Gamma = 2$, and defining $\epsilon = 1 - a_\infty^2 \ll 1$, the coefficients (\ref{cubic_components}) reduce to
\begin{equation}
    A = - 5/3, ~~~~B = 1/3, ~~~~C = (1 - \epsilon^2)/3
    \hfill (\Gamma = 2)
\end{equation}
and we can factor the cubic equation (\ref{gen_cubic}), to leading order, as
\begin{equation}
    (x - x_1)(x - x_2)(x - x_3) = {\mathcal O}(\epsilon^6)
\end{equation}
with 
\begin{equation} \label{as_stiff_highdensity}
\begin{split}
    x_1 & = -1/3 + 3 \epsilon^2 / 16 + 27 \epsilon^4 /512 ,\\
    x_2 & = 1 + \epsilon/2 - 3 \epsilon^2 / 32 
    + 90 \epsilon^3 / 2048 - 27 \epsilon^4 / 1024, \\
    x_3 & = 1 - \epsilon/2 - 3 \epsilon^2 / 32
    - 90 \epsilon^3 / 2048 - 27 \epsilon^4 / 1024,
    \end{split}
    \hfill (\Gamma = 2)
\end{equation}
(cf.~Fig.~\ref{fig:cubic_solutions}).  Since only $x_3$ satisfies $0 \leq x_3 \leq 1$, we identify this solution with our physical solution.  Inserting this together with $a_\infty^2 = 1 - \epsilon$ into (\ref{lambda_GR}) we can expand the accretion rate eigenvalue $\lambda_{\rm GR}$ about $\epsilon = 0$ to find, 
\begin{equation} \label{lambda2}
\begin{split}
    \lambda_{\rm GR} & \simeq 4 - 4 \epsilon + \frac{5}{16} \epsilon^2 \\
    & = 4 a_\infty^2 + \frac{5}{16} (1 - a_\infty^2)^2
    \end{split}
    \hfill (1 - a_\infty^2 \ll 1,~\Gamma = 2)
\end{equation}

Now focus on the ``ultra-relativistic" EOS
\begin{equation} \label{ur}
P = \rho,
\end{equation}
for which the sound speed exactly equals the speed of light, i.e.~$a = 1$ (see \eq~\ref{soundspeed}). \eq~(\ref{ur}) is a special case of a $\Gamma$-law equation of state with $\Gamma = 2$ in the limit of $\rho \gg \rho_0$.  It is sometimes invoked as the extreme (causal) limit of possible EOSs governing the core of neutron stars, and has therefore been used in establishing various limits on the maximum mass of neutron stars \citep[see, e.g.][]{RhoR74,KorSF97,Bayetal18,RuiST18}.  Using $a_\infty^2 = 1$ in (\ref{lambda2}) we have $\lambda_{\rm GR} = 4$, and \eq~(\ref{Mdot}) reduces to
\begin{equation}
\dot M = 16 \pi M^2\rho_{0\infty}. \hfill (P = \rho)
\label{spher}
\end{equation}
In fact, \cite{PetST88} found the surprising analytic result that for a (rotating) Kerr black hole, moving with 3-velocity $\varv_{\infty}$ in a $P=\rho$ gas that is uniform and at rest far from the black hole, the steady-state accretion rate does not depend on the orientation of the spin or the direction of the flow! The rate is given exactly by 
\begin{equation}
\dot M = 4 \pi (r_+^2 + a_{\rm h}^2) \, \gamma_{\infty} \rho_{0\infty},
\label{Petrich}
\end{equation}
where $r_+ = M+(M^2 - a_{\rm h}^2)^{1/2}$ is the radius of the event horizon, $a_h=J_{\rm h}/M$ is the spin parameter of the black hole with angular momentum  $J_{\rm h}$ and $\gamma_\infty = 1/(1-\varv_{\infty}^2)^{1/2}$. For a non-spinning black hole at rest ($a_{\rm h}=0=\varv_{\infty}$, $\gamma_{\infty} = 1$ and $r_+ = 2M$), \eq~(\ref{Petrich}) reduces to the spherical accretion value given by
\eq~(\ref{spher}), as expected.

\section{Summary}

To summarize, a fully relativistic treatment is always necessary to describe spherical accretion {\it flow} (e.g., the density and velocity profiles) onto a Schwarzschild black  hole in the vicinity of the event horizon, where nonlinear gravity is important. However, when $a^2_{\infty} \ll 1$ and $ 1 \le \Gamma \le 5/3$ the steady-state accretion {\it rate} is given reliably by the Newtonian Bondi result, \eq~(\ref{Mdot}) with $\lambda_{\rm GR} = \lambda_s$, given by (\ref{lambda_s}).  This is a consequence of the fact that the rate is determined by conditions at the transonic radius,  which is far outside the horizon ($r_s \gg M$) where nonlinear gravity is unimportant.  This conclusion  holds even in the case of $\Gamma = 5/3$, although determining the transonic radius for $\Gamma = 5/3$ requires the relativistic solution (ST, Ex.~G.1)! By contrast, for $a^2_{\infty } \lesssim 1$, or for $\Gamma > 5/3$, a relativistic solution is necessary to describe both the accretion flow {\it and} the accretion rate. 

In particular, we complement earlier treatments of Bondi accretion for stiff equations of state, and provide relativistic expressions for the eigenvalue $\lambda_{\rm GR}$ which determines the accretion rate.  We discuss qualitative differences between accretion for soft and stiff EOSs, including the  appearance of a minimum (steady-state) accretion rate for $\Gamma \geq 5/3$, and an associated asymptotic power-low density profile with $\rho_{0} \rightarrow 0$ as $r \rightarrow 0$.  We explore these qualitative difference by considering, as limiting cases, both low and high-density regimes, providing closed-form expressions for many of the leading-order terms.  As a special example we consider an EOS with $\Gamma = 2$, which is particularly useful for modeling neutron stars supported by a stiff EOS, as well as its ultra-relativistic limiting case $P = \rho$.  This EOS provides a natural limit for the unknown cold, nuclear EOS above nuclear density.  In addition to being interesting in their own right, our results are useful for estimates of accretion rates onto, for example, mini-black holes residing at the core of neutron stars, and hence for determining the lifetimes of such objects.  We will use this rate to diagnose our numerical simulations of such a scenario in a forthcoming paper \citep{RicBS21b}.

\section*{Acknowledgments}

CBR acknowledges support through an undergraduate research fellowship at Bowdoin College, and would like to thank Maria Perez Mendoza for many helpful conversations.  This work was supported in parts by National Science Foundation (NSF) grants PHY-2010394 to Bowdoin College, and NSF grants  PHY-1662211 and PHY-2006066 and NASA grant 80NSSC17K0070 to the University of Illinois at Urbana-Champaign.

\section*{Data Availability}

There are no new data associated with this article.

\bibliographystyle{mnras}

\begin{appendix}
\section{Accretion Profiles}
\label{sec:profiles}

\subsection{Structure equations}
\label{sec:profiles:equations}

Assuming stationary and isentropic fluid flow in spherical symmetry, the relativistic Euler equation together with the equation of baryon conservation can be combined to yield
\begin{equation} \label{prime}
\frac{d u}{dr} = \frac{D_1}{D},~~~~~~~~~~~~\frac{d \rho_0}{dr} = - \frac{D_2}{D},
\end{equation}
where we have used Schwarzschild coordinates, and where the coefficients $D_1$, $D_2$ and $D$ are given by
\begin{equation} \label{flow_coefficients}
    \begin{split}
        D_1 & = \frac{1}{\rho_0} \left\{ \left(1 - \frac{2M}{r} + u^2 \right) \frac{2 a^2}{r} - \frac{M}{r^2} \right\} \\
        D_2 & = \frac{1}{u} \left( \frac{2 u^2}{r} - \frac{M}{r^2} \right) \\
        D & = \frac{1}{\rho_0 u} \left\{ u^2 - \left(1 - \frac{2M}{r} + u^2\right) a^2 \right\}
    \end{split}
\end{equation}
(see, e.g., ST G.10 -- ST G.13).  The critical radius $r_s$ is defined by the simultaneous vanishing of these three coefficients, which yields the conditions (\ref{u_s}) and (\ref{r_s}).  The rest-mass density $\rho_0$ and the (inward) fluid velocity $u$ are also related by
\begin{equation} \label{Mdot_general}
4 \pi r^2 \rho_0 u = const = \dot M,
\end{equation}
where $\dot M$ is the accretion rate.

Once values for the critical radius have been determined, \eqs~(\ref{prime}) can be integrated from the critical point toward both smaller and larger radii, resulting in profiles of the fluid density and velocity (see Figs.~\ref{fig:profile4o3_large_a} through \ref{fig:profile2} for examples).

\subsection{Asymptotic behavior}
\label{sec:profiles:asymptotics}

It is also instructive to consider the asymptotic behavior of solutions to \eqs~(\ref{prime}).  We will assume $a^2_\infty \ll 1$, so that we may adopt the leading-order relation (\ref{density_small_a}) to express the sound speed $a$ in terms of the rest-mass density $\rho_0$.  Abbreviating $\alpha = \dot M / (4 \pi)$ we may also write
\begin{equation} \label{u_of_alpha}
u = \frac{\alpha}{\rho_0 r^2}.
\end{equation}
Using these relations, we can write the second equation in (\ref{prime}) as an equation for $\rho_0$ alone,
\begin{equation} \label{rho_prime}
\frac{d \rho_0}{dr} = - \frac{\rho_0}{r} \, \frac{2 \alpha^2 - M \rho_0^2 r^3}{\alpha^2 - (\rho_0^2 r^4 - 2 M \rho_0^2 r^3 + \alpha^2) \,\Gamma K \rho_0^{\Gamma - 1} }.
\end{equation}
The asymptotic behavior will depend on which terms in the last equation dominate at large radii.  Assuming that 
\begin{equation} \label{assumption}
M \rho_0^2 r^3 \gg \alpha^2, 
\end{equation}
which will certainly be the case if $\rho_0$ approaches a non-zero asymptotic value $\rho_{0\infty}$, we may approximate (\ref{rho_prime}) as
\begin{equation}
\frac{d \rho_0}{dr} = - \frac{M}{\Gamma K} \, \frac{\rho_0^{2 - \Gamma}}{r^2}. \end{equation}
Integrating this equation then yields the asymptotic density profile
\begin{equation} \label{asymptotic_profile}
    \rho_0 \simeq \left(\frac{\Gamma - 1}{\Gamma} \frac{M}{K} \frac{1}{r} + \rho_{0\infty}^{\Gamma - 1} \right)^{1/(\Gamma - 1)}.
\end{equation}
Given the accretion rate $\dot M$, as determined earlier, the profile for the four-velocity can then be found from (\ref{u_of_alpha}).  The approximate, asymptotic profile (\ref{asymptotic_profile}), and corresponding values for $u$, are included as the faint lines in Figs.~\ref{fig:profile4o3} and \ref{fig:profile2}.

We now consider, as a limiting case, the limit $\rho_{0\infty} \rightarrow 0$, in which case the solution (\ref{asymptotic_profile}) reduces to
\begin{equation} \label{powerlaw_profile}
    \rho_0 \simeq \left(\frac{\Gamma - 1}{\Gamma} \frac{M}{K} \frac{1}{r} \right)^{1/(\Gamma - 1)}.
\end{equation}
We now observe that the condition (\ref{assumption}) is still satisfied for $\Gamma > 5/3$, but not for $\Gamma < 5/3$.  Therefore, the power-law profile (\ref{powerlaw_profile}) describes the asymptotic behavior of a non-vanishing accretion solution with zero density at infinity -- but only for $\Gamma > 5/3$.  It is this solution that corresponds to the minimum accretion rate that we encountered in Section \ref{sec:small_a_large_gamma}.

We may also consider the opposite assumption, namely that
\begin{equation} \label{opposite_assumption}
M \rho_0^2 r^3 \ll \alpha^2 
\end{equation}
(which evidently requires that $\rho_{0\infty} = 0$).  Further assuming that $\rho_0$ is monotonically decreasing, i.e.~$d \rho_0 / dr < 0$, \eq~(\ref{rho_prime}) reduces to
\begin{equation}
    \frac{d \rho_0}{dr} = - \frac{2 \rho_0}{r},
\end{equation}
which is solved by $\rho_0 = (\kappa / r)^2$, where $\kappa$ is a constant of integration.  In this case \eq~(\ref{u_of_alpha}) yields $u \rightarrow 4 \pi \dot M / \kappa^2$, contradicting our assumption that $u \rightarrow 0$ asymptotically.  We therefore conclude that non-trivial solutions with $\rho_{0\infty} = 0$ exist only for $\Gamma > 5/3$, in which case they are given by (\ref{powerlaw_profile}).

\subsection{Transonic behavior}
\label{sec:transonic}

For completeness, we note that for transonic flow deep inside the critical radius $r_s$ the velocity approaches free-fall,
\begin{equation} \label{u_freefall}
u^2  \approx \frac{2M}{r}, \hfill (u \gg a,~r \ll r_s)  
\end{equation}
in which case the density becomes
\begin{equation} \label{rho0_freefall}
\rho_0 = \frac{\dot M}{4 \pi u r^2}  
\approx  \frac{\dot M}{2^{5/2} \pi  M}\,  \frac{1}{r^{3/2}}.
\hfill (u \gg a,~r \ll r_s)
\end{equation}
If $r_s \gg 2 M$, the fluid flow just outside the black hole horizon, where typically the gas makes its greatest contributions to any perturbative outgoing radiation, is well approximated by the above expressions \citep[][see also ST for discussion and references]{Sha73a}.  The scaling of $u$ and $\rho_0$ with $r$ given by (\ref{u_freefall}) and (\ref{rho0_freefall}) is evident in Figs.~\ref{fig:profile4o3_large_a} through \ref{fig:profile4o3} and even applies for $u \sim a$ inside the transonic radius if $\Gamma = 5/3$. 

\section{Double roots}
\label{sec:double_roots}

In this brief appendix we derive the double roots (\ref{double_roots}).  We start by factoring out equation (\ref{double_root_factors}) and identifying the resulting coefficients with those of (\ref{gen_cubic}), which results in
\begin{equation} \label{app_double_root_coefficients}
    \begin{split}
        - (2 x_1 + x_2) & = A = (7 - 6 \Gamma)/3  \\
        2 x_1 x_2 + x_1^2 & = B = (1 - \Gamma)(5 - 3 \Gamma)/3\\
        - x_1^2 x_2 & = C = a_\infty^2 \, (2 \Gamma - 2 - a_\infty^2) / 3.
    \end{split}
\end{equation}
Using the first of these equations to eliminate $x_2$ in the second results in a quadratic equation for $x_1$ with the two solutions
\begin{equation} \label{app_x1}
    (x_1)_{1/2} = \left\{
\begin{array}{l}
(3 \Gamma - 5)/9 \\
\Gamma - 1.
\end{array}
\right.
\end{equation}
From the first equation in (\ref{app_double_root_coefficients}) we then have
\begin{equation} \label{app_x2}
    (x_2)_{1/2} = \left\{
\begin{array}{l}
(12 \Gamma - 11)/9 \\
-1/3.
\end{array}
\right.
\end{equation}
We now insert (\ref{app_x1}) and (\ref{app_x2}) into the third equation of (\ref{app_double_root_coefficients}), which yields a quadratic equation for $a_\infty^2$.  For the top solutions in (\ref{app_x1}) and (\ref{app_x2}) the discriminant is negative, so that we have no real solutions.  For the bottom solutions the discriminant vanishes, so that we obtain
\begin{equation}
    a_\infty^2 = \Gamma - 1
\end{equation}
as the only viable solution.  Since $x_2 = -1/3$ is negative for this solution, we identify $a_s^2$ with $x_1$ and obtain (\ref{double_roots}).

\end{appendix}

\label{lastpage}
\end{document}